\documentstyle[psfig]{l-aa}
\def\la{\;
\raise0.3ex\hbox{$<$\kern-0.75em\raise-1.1ex\hbox{$\sim$}}\; }
\def\ga{\;
\raise0.3ex\hbox{$>$\kern-0.75em\raise-1.1ex\hbox{$\sim$}}\; }
\begin{document}
\thesaurus{11(02.12.1; 02.12.3; 11.17.1; 11.17.4 APM 08279+5255)}
\title{Monte Carlo inversion of hydrogen and metal lines
from QSO absorption spectra
\thanks{Based in part on data 
obtained at the W. M. Keck Observatory, which is jointly operated by
the California Institute of Technology, the University of
California and the National Aeronautics and Space Administration.}
}
\author{Sergei A. Levshakov\inst{1}
\and Irina I. Agafonova\inst{1}
\and Wilhelm H. Kegel\inst{2}}
\offprints{S.~A.~Levshakov }
\institute{
Department of Theoretical Astrophysics,
A. F. Ioffe Physico-Technical Institute, 194021 St. Petersburg, Russia
\and
Institut f\"ur Theoretische Physik der Universit\"at Frankfurt am Main,
60054 Frankfurt/Main 11, Germany}
\date{Received March 00, 2000; accepted March 00, 2000}
\maketitle
\markboth{S.A. Levshakov et al.: Monte Carlo inversion of line profiles}{ }
%\maintitlerunninghead{}
%\authorrunninghead{S.A. Levshakov et al.}
%
\begin{abstract}
A new method, based on the simulated annealing algorithm and aimed at the inverse
problem in the analysis of intergalactic (interstellar) complex spectra of
hydrogen and metal lines, is presented. 
We consider the process of line formation in clumpy stochastic media accounting
for fluctuating velocity and density fields 
(mesoturbulence).
This approach generalizes our previous Reverse Monte Carlo 
and Entropy-Regularized Minimization methods which were 
applied to velocity fluctuations only. 
The method allows one to estimate, from an observed system of spectral lines,
both the physical parameters of the absorbing gas and appropriate structures
of the velocity and density distributions along the line of sight. The validity
of the computational procedure is demonstrated using a series of synthetic spectra
that emulate the up-to-date best quality data. 
H\,{\sc i}, C\,{\sc ii}, Si\,{\sc ii}, C\,{\sc iv},
Si\,{\sc iv}, and O\,{\sc vi} lines, exhibiting complex profiles,
were fitted simultaneously. The adopted physical parameters
have been recovered with a sufficiently high accuracy.
The results obtained encourage the application of the proposed
procedure to the analysis of real observational data.

\end{abstract}

\section{Introduction}

QSO absorption line spectroscopy being a major activity at many
observatories for the last two decades is now developing into
a powerful tool for extragalactic research thanks to the new
generation of large telescopes. 
The steady improvement in sensitivity
and resolution of spectroscopic instrumentation opens new fields
in the study of QSO absorption systems. 
It is now becoming possible to investigate the intensity fluctuations
within the line profiles and thus to estimate hydrodynamic
characteristics of the absorbing gas.

The majority of the narrow QSO
absorption lines represents intervening systems
and allows us to probe the properties of diffuse matter
at very high redshifts. Resolved profiles of hydrogen lines and 
especially lines of heavier elements (`metals') show a diversity
of shapes and structures. 
Up to now, their analysis is based on
the assumption that the observed complexity is
caused by individual `clouds' randomly distributed
along the line of sight with slightly different radial velocities. 
It is also a basic assumption that the hydrodynamic (`bulk' or `turbulent')
velocity distribution inside each cloud is Gaussian and
completely uncorrelated ({\it microturbulence}).
This model implies that each subcomponent of the complex profile being
resolved should have a symmetrical profile and its shape should not
alter with higher spectral resolution. 
Observations show, however,
that the complexity of the line profiles increases with higher resolution,
a tendency expected for correlated bulk motions 
which have, in general, non-Gaussian distributions along a given line of sight
(Levshakov \& Kegel 1997; Levshakov, Kegel \& Mazets 1997;
Levshakov, Kegel \& Takahara 1999; Papers~I, II, and III
hereafter, respectively).
It follows that the microturbulent approximation is not appropriate 
in this case because it does not
account for all the relevant physical processes involved
in the radiative transfer. Moreover, being applied to real data,
the microturbulent analysis leads to a loss of valuable information
contained in the observations and may even yield unphysical results
(Levshakov \& Kegel 1999;
Levshakov, Takahara \& Agafonova 1999; LTA hereafter).
The need for more sophisticated procedures of data analysis becomes
therefore obvious.

In recent years, it has been shown that accounting for the correlations
in the velocity field ({\it mesoturbulence}) may change the interpretation
of the line measurements substantially (Papers~I and II).
A mesoturbulent approach has been already
successfully applied to the study of the deuterium and hydrogen absorption
in Q~1937--1009 (Levshakov, Kegel \& Takahara 1998a), Q~1718+4807
(Levshakov, Kegel \& Takahara 1998b), and Q~1009+2956
(Levshakov, Tytler \& Burles 2000).
For all three QSOs about the same value for the D/H ratio
was derived in contrast to the 
previously announced microturbulent results.
Our first inversion codes, --  the Reverse Monte Carlo 
(Paper~III), and the Entropy-Regularized Minimization
(LTA), -- have been developed for a model of 
a stochastic velocity field neglecting any density fluctuations. They
have been applied to the analysis of the H\,{\sc i}
and D\,{\sc i} lines and/or to the metal absorption lines with
similar profiles when species trace the same volume elements
independently on the density fluctuations. 
In the present paper, we extend this study to the inverse
problem for a model of compressible turbulence when one observes
non-similar profiles of different atoms and/or ions  
from the same absorption-line system.
As in our previous papers, we use the term `turbulence' in a wider
sense as compared with hydrodynamic turbulence to label the
unknown nature of the line broadening mechanism.
In this regard
we consider any kind of  
bulk motions (infall, outflows, tidal flows etc.)
of more or less stochastic nature
leading to
fluctuating velocity and density (temperature) fields 
as continuous random functions of
the space coordinate along a given line of sight
within the intervening absorbing region.

Two noteworthy works have been recently carried out aiming at 
the recovery of the
physical intergalactic structure from the Lyman-$\alpha$ forest lines. 
Nusser \& Haehnelt (1999a,b)
developed an inverse procedure based 
on the relation between density and velocity Fourier coefficients.
The quality of their
recovery is, however, restricted by
the assumption that the Lyman-$\alpha$ forest structure traces 
mainly the matter density distribution and that the
amplitude of the peculiar velocities is rather small to
affect the local absorption coefficient significantly.
This assumption is questionable since
there is no simple way to distinguish {\it observationally} whether
the density or the velocity fluctuations are the main source of the
`line-like' structure observed in the
Lyman-$\alpha$ forest (Levshakov \& Kegel 1998).
Moreover, recent studies of nearby large-scale motions
in the universe indicate that the Hubble flow is considerably
perturbed. 
Peculiar velocities in the range from 300 to 500 km~s$^{-1}$
have been found in a sample of galaxies complete out to
a distance of 60~Mpc (e.g., Watkins 1997; 
Gramann 1998; Giovanelli et al. 1998), 
a fact which should be taken into account  
in the inverse procedures.

The method described in the present paper is quite
flexible and equally accounts for the density and velocity fluctuations.
It is based on a stochastic optimization
approach similar to that developed in Paper~III.
We estimate simultaneously the 
physical parameters {\it and} appropriate realizations of
the density $n(s)$ and velocity $v(s)$ distributions
along the line of sight
to reproduce hydrogen and metal lines from a 
given absorption system. 
In this regard, the more spectra of different elements are incorporated
in the analysis the higher accuracy of the estimation can be obtained.
 
In $\S 2$ our model and the underlying basic assumptions are specified.
The inversion code is described in $\S 3$.
The validity of the method is tested in $\S 4$ using
simulated sets of noisy line profiles
(H\,{\sc i}, C\,{\sc ii}, Si\,{\sc ii}, C\,{\sc iv},
Si\,{\sc iv}, and O\,{\sc vi}).
Finally, the main conclusions are outlined in $\S 5$.

\section{Hydrogen and metal absorption  
from QSO Lyman-limit systems}

In this section we consider the line formation in a
Lyman-limit system (LLS), -- the
intervening absorbing gas being optically thin in the Lyman continuum
(presumably outer regions of a foreground galaxy).
To specify the calculations, we use the standard
photoionization model of Donahue \& Shull (1991, hereafter DS), namely,
an extended region of thickness $L$ with a given metallicity. 
The region is ionized by a background
photoionizing spectrum given by Mathews \& Ferland (1987).
The gas is assumed to be in thermal equilibrium.

We concentrate our efforts on the LLSs because, on one hand, they can be 
analyzed with a minimum number of model assumptions and, on the
other hand, they 
are the most promising targets for measuring 
the extragalactic deuterium to hydrogen
ratio (Burles \& Tytler 1998a,b). 
In addition, one
may expect that kinematic characteristics such as 
dispersions of the velocity and density fluctuations within LLSs
are directly related to the processes of galaxy formation.
If this is true, these characteristics should change with
cosmic time (i.e. with $z$), and we can estimate them through
the inversion procedure in question.

The LLSs often show carbon and silicon line absorption from
different ionization stages and even  O\,{\sc vi} lines
(e.g., Kirkman \& Tytler 1999). 
The electron density in the LLSs is rather low,
$n_{\rm e} \sim 10^{-2} - 10^{-3}$~cm$^{-3}$, and the kinetic
temperature $T$ is about $10^4$~K. 
Under these conditions photoionization dominates the ionization
structure and the relative abundance 
of different ions of the same element is a function 
of the density only, once the radiation field has been
specified.

With the assumed spectral distribution
it is convenient to describe the thermal and the ionization state of the
gas in the LLSs through the dimensionless `ionization parameter'
$U = n_{\rm ph}/n_{\rm H}$, equal to the ratio of the number
of photons with energies above one Rydberg to the 
total hydrogen density.
Following DS,
we assume that the ionizing background radiation is not
time or space dependent.
Then for a given value of $n_{\rm ph}$ (or the specific
radiation flux $J_0$ at 1~Ry), the distribution of $U$ along
the
line of sight represents the reciprocal hydrogen density.

Within the absorbing region the velocity distribution $v(s)$ for 
all species is the same, but their fractional ionizations  may vary
significantly along the line of sight and therefore their apparent
profiles may show a diversity of shapes. This allows us
to tackle the inverse problem, i.e. to find such velocity,
gas density and temperature distributions along the line of sight that 
provide the observed variety of the line profiles.   

\subsection{Basic equations}

We consider the formation of absorption lines in  the light
of a point-like source of continuum radiation, i.e.
the absorbing region is assumed to be far from the quasar.
The observed 
absorption spectra therefore correspond to only one line of sight,
i.e. to one realization of the radial velocity and gas density
distributions. 

For a resolved and optically thin absorption line,
one observes directly the apparent optical depth
$\tau^\ast(\lambda)$ as function of wavelength $\lambda$
\begin{equation}
\tau^\ast(\lambda) = 
\ln \left( I_{\rm c}/I_\lambda\right)\: ,
\label{eq:E1}
\end{equation}
where $I_\lambda$ and $I_{\rm c}$ are the 
intensities in the line and in the continuum, respectively.

The recorded spectrum is a convolution of the true spectrum and the
spectrograph point-spread function $\phi_{\rm sp}$
\begin{equation}
I_\lambda = \int\,I_{\rm c}\,
{\rm e}^{-\tau(\lambda')}\,
\phi_{\rm sp}(\lambda - \lambda')\,{\rm d}\lambda'\: ,
\label{eq:E2}
\end{equation}
where $\tau(\lambda)$ is the true (intrinsic) optical depth.

The intensity of the background continuum source $I_{\rm c}$
changes usually very slowly over the width of the spectrograph function,
and hence the normalized observed intensity is
\begin{equation}
{\cal F}_\lambda \equiv
\frac{I_\lambda}{I_{\rm c}} = \int\,
{\rm e}^{-\tau(\lambda')}\,
\phi_{\rm sp}(\lambda - \lambda')\,{\rm d}\lambda'\: .
\label{eq:E3}
\end{equation}

The instrumental point-spread function $\phi_{\rm sp}$ can be determined
experimentally, whereas $\tau(\lambda)$ is
the quantity of astrophysical interest.
The true optical depth is defined through the local absorption
coefficient $\kappa_\lambda(s)$ by
\begin{equation}
\tau(\lambda) =
\int^{L}_{0}\,\kappa_\lambda(s)\,{\rm d}s = 
L\int^1_0\,\kappa_\lambda(x)\,{\rm d}x\; ,
\label{eq:E4}
\end{equation}
where $x = s/L$ is the normalized coordinate along the line of sight.

In this equation $\kappa_\lambda(x)$ is a stochastic variable which
depends on the random realization of three fields~: velocity,
density and temperature. We write $\kappa_\lambda$
as the product of the local absorption cross-section 
per atom, $k_\lambda(x)$, and the local number density of
absorbing atoms, $n(x)$
\begin{equation}
\kappa_\lambda(x) = k_\lambda(x)\,n(x)\:  .
\label{eq:E5}
\end{equation}

At point $x$, the absorption cross-section has the form
\begin{equation}
k_\lambda(x) = k_0\,\Phi_\lambda[\Delta\lambda_{\rm D}(x),v(x)]\: .
\label{eq:E6}
\end{equation}

The quantity $k_0$ is a constant for a particular line
\begin{equation}
k_0 = \frac{\pi e^2}{m_{\rm e} c^2}f_{\rm abs}\lambda^2_0\, ,
\label{eq:E7}
\end{equation}
where $m_{\rm e}$ and $e$ are the mass and charge of the electron,
$f_{\rm abs}$ is the oscillator strength of the line for
absorption, and $\lambda_0$ is the rest wavelength of the line
center.

The profile function $\Phi$
describes the local broadening at point $x$.
It is Doppler shifted according to the local velocity
$v(x)$. Thus we have
\begin{equation}
\Phi_\lambda[\Delta\lambda_{\rm D}(x),v(x)] =
\Phi_\lambda\{\Delta\lambda_{\rm D}(x),[\lambda - \lambda_0 -
\lambda_0\frac{v(x)}{c}]\}\: .
\label{eq:E8}
\end{equation}

Under interstellar/intergalactic conditions the profile function
$\Phi$ is determined in general by the Doppler effect and by radiation
damping, i.e. it corresponds to a Voigt profile.
The Doppler width of the line, $\Delta\lambda_{\rm D}$,
is determined in this case by the thermal width, $v_{\rm th}$,
\begin{equation}
\Delta\lambda_{\rm D}(x) = \lambda_0 \frac{v_{\rm th}(x)}{c} =
\frac{\lambda_0}{c}\,\sqrt{2 k_{\rm B} T(x)/m_{\rm a}}\: ,
\label{eq:E9}
\end{equation}
and hence $v_{\rm th}$ characterizes the width of the
local absorption coefficient at a given value of $v$ (here
$k_{\rm B}$ is Boltzmann's constant, $m_{\rm a}$ is 
the mass of the ion under consideration,
and $T(x)$ is the local kinetic temperature).

Now, let $n_{\rm a}(x)$ be the local number density
of element `a', $n_{{\rm a},i}(x)$ be the density of ions
in the $i$th ionization stage, and 
$Z_{\rm a} = n_{\rm a}/n_{\rm H}$ be the relative abundance
of this element which is regarded 
to be a constant along the line of sight.
According to the standard model of DS,
the fractional ionization of ions \{${\rm a},i$\}
\begin{equation}
\Upsilon_{{\rm a},i} = \frac{n_{{\rm a},i}(x)}{n_{\rm a}(x)}
\label{eq:E10}
\end{equation}
may be described by a function of $U$ only, 
$\Upsilon_{{\rm a},i} = \Upsilon_{{\rm a},i}[U(x)]$. 

Then for a fixed
value of $\lambda$ within the line profile, the optical
depth is given according to equations (\ref{eq:E4})--(\ref{eq:E6}) by
\begin{eqnarray}
\lefteqn{
\tau_{{\rm a},i}(\lambda)\, = } \nonumber \\
& & k_0 L \int^1_0\,
n_{\rm a}(x)\,\Upsilon_{{\rm a},i}[U(x)]\,
\Phi_\lambda[\Delta\lambda_{\rm D}(x),v(x)]\,{\rm d}x ,
\label{eq:E11}
\end{eqnarray}
or with $n_{\rm a}(x) = Z_{\rm a} n_{\rm H}(x)$ 
\begin{eqnarray}
\lefteqn{
\tau_{{\rm a},i}(\lambda)\, = } \nonumber \\
& & k_0 Z_{\rm a} L \int^1_0\,
n_{\rm H}(x)\,\Upsilon_{{\rm a},i}[U(x)]\,
\Phi_\lambda[\Delta\lambda_{\rm D}(x),v(x)]\,{\rm d}x .
\label{eq:E12}
\end{eqnarray}

\begin{figure}
\vspace{5.0cm}
\hspace{-0.7cm}\psfig{figure=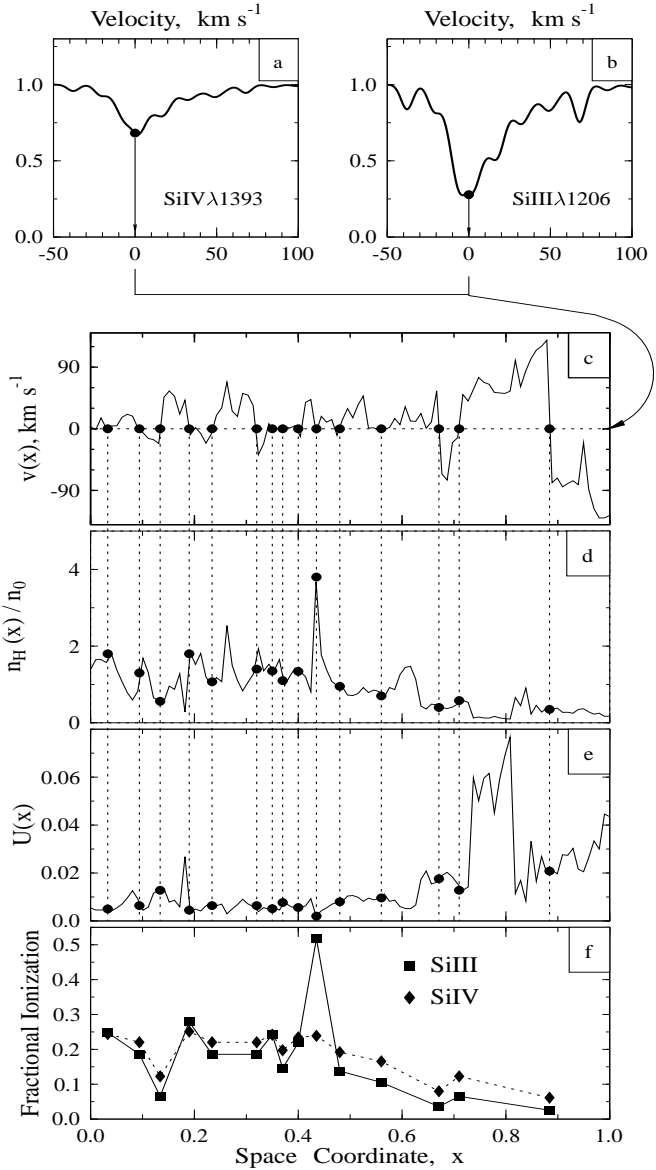,height=12.0cm,width=8.0cm}
\vspace{-1.5cm}
\caption[]{Velocity plots of Si\,{\sc iv} (a) and Si\,{\sc iii} 
(b) synthetic profiles 
calculated for the $z_{\rm a} = 3.514$ system towards QSO~08279+5255 by Levshakov,
Agafonova \& Kegel (2000). The underlying velocity, density and ionization
parameter distributions are shown in panels (c), (d), and (e), respectively.
The diversity of fractional ionizations in the marked areas is 
illustrated in panel (f) }
\end{figure}

Having defined $n_{\rm H}(x)$, 
the ionization fractions $\Upsilon_{{\rm a},i}$ can be calculated
using a photoionization model corresponding to that of DS or Ferland (1995).
In the present paper we used the results of DS since we may neglect
the radiation transfer and high-density effects incorporated in 
Ferland's CLOUDY code.

\subsection{Some aspects of line formation in clumpy 
stochastic media}

When using the conventional Voigt-profile fitting (VPF) procedure
several subcomponents are usually assumed 
to describe an absorption line which is too complex 
to be fitted by  a single Voigt profile.
Each component is supposed to represent a homogeneous plane-parallel
slab of gas with its own physical parameters~: radial velocity, the
rms velocity, density, ionization and 
element abundances (see e.g. Tytler et al. 1996).
This fitting method would yield physically reasonable results for gas
in spatially separate homogeneous clouds if we could know {\it a priori}
the actual number of these clouds and have a justification for their
homogeneity. As far as the number of subcomponents is unknown, 
the VPF method loses uniqueness and stability.
The main question remains whether the complex line shape is caused by
separate homogeneous clouds or by a continuous medium with fluctuating
characteristics. There are no unambiguous observational data favoring one 
or the other assumption. But some indirect information such as the increasing
complexity of the line profiles with increasing spectral resolution and
recently performed 3D hydrodynamic calculations 
(see e.g. Rauch, Haehnelt \& Steinmetz 1997)
give higher weight to the model of a fluctuating continuous medium.
Applied to the inversion problem, this
has important consequences.

Fig.~1 illustrates step-by-step the line formation process in a
stochastic medium. The synthetic spectra of Si\,{\sc iv}$\lambda1393$
(panel {\bf a})
and Si\,{\sc iii}$\lambda1206$ (panel {\bf b})
are formed in an absorbing region with the fluctuating velocity
and density fields shown in panels {\bf c} and {\bf d}, respectively. 
The corresponding distribution of the ionization parameter $U(x)$ is
shown in panel {\bf e}. 

Let us consider a point within the
Si\,{\sc iv} and Si\,{\sc iii} profiles, for example at the radial
velocity $v = 0$~km~s$^{-1}$ (panels {\bf a} and {\bf b}). 
Then filled circles in panel {\bf c} mark those volume elements 
having the same radial velocity. The corresponding points in panel {\bf d}
show the density in the corresponding 
areas contributing to the `observed' intensity.

It is clearly seen that for a given point within the line profile
the observed intensity results from a mixture of different ionization states
(labeled portions of the $U$-distribution in panel {\bf e}). It is also
obvious that each point of the line profile is caused by a
different combination of these states. 
As an example, panel {\bf f} shows  the different fractional ionizations 
of Si\,{\sc iii} and Si\,{\sc iv} at the points with $v = 0$ km~s$^{-1}$. 

In practical applications, the metal abundances, $Z_{\rm a}$, 
are usually estimated 
from the ratio of the total column densities $N_{\rm a}$ and $N_{\rm H}$
by means of the `correction for the ionization'~:
\begin{equation}
Z_{\rm a} = \frac{N_{{\rm a},i}}{N_{{\rm H}\,{\sc i}}}
\,\frac{\Upsilon_{{\rm H}\,{\sc i}}(\bar{U})}{\Upsilon_{{\rm a},i}(\bar{U})}\: .
\label{eq:E24}
\end{equation}

The mean value of $\bar{U}$ is usually estimated from the ratio 
$N_{{\rm a},i}/N_{{\rm a},j}$.
While this procedure is suggestive, 
it is {\it incorrect} if the degree of ionization
varies along the line of sight. 
From eq.~(\ref{eq:E12}) and accounting for 
$\int\,\Phi_\lambda\,{\rm d}\lambda  = 1$, it follows that
\begin{equation}
\frac{N_{{\rm a},i}}{N_{{\rm a},j}} =
\frac{\int\,n_{\rm H}(x)\,\Upsilon_{{\rm a},i}[U(x)]\,{\rm d}x} 
{\int\,n_{\rm H}(x)\,\Upsilon_{{\rm a},j}[U(x)]\,{\rm d}x} =
\frac{\bar{\Upsilon}_{{\rm a},i}}{\bar{\Upsilon}_{{\rm a},j}}\: ,
\label{eq:E25}
\end{equation}
where $\bar{\Upsilon}_{{\rm a},i}$ and $\bar{\Upsilon}_{{\rm a},j}$
denote the mean density-weighted fractional ionizations. 

The ratio $\Upsilon_{{\rm a},i}/\Upsilon_{{\rm a},j}$
as a function of $U$ can be calculated from the adopted photoionization
model. Then the mean value of the ionization parameter, $\bar{U}$,
is estimated using the implicitly assumed relation
\begin{equation}
\frac{\bar{\Upsilon}_{{\rm a},i}}{\bar{\Upsilon}_{{\rm a},j}} =
\frac{\Upsilon_{{\rm a},i}(\bar{U})}{\Upsilon_{{\rm a},j}(\bar{U})}\: .
\label{eq:E25a}
\end{equation}

Since $\Upsilon_{{\rm a},i}$ is a non-linear function of $n_{\rm H}$, 
it is clear that the equality $\bar{\Upsilon} = \Upsilon(\bar{U})$
holds only in the case of a uniform density. Applying relation (\ref{eq:E25a}) 
to the case of a
fluctuating gas density leads to a wrong value of $\bar{U}$ --
and moreover the estimation of the mean ionization parameter will
depend on the specific pair of ions chosen to calculate the
left-hand side of (\ref{eq:E25}).

To illustrate the numerical differences that may occur between the micro- and 
mesoturbulent approaches we consider 
the following example. 
The absorption lines shown in Fig.~1
have approximately the same total column
densities [the data are taken from
Ellison et al. 1999 ($\log N_{{\rm Si}\,{\sc iv}} \simeq 12.8$~cm$^{-2}$)
and Molaro et al. 1999
($\log N_{{\rm Si}\,{\sc iii}} \simeq 12.8$~cm$^{-2}$)].
The ratio $N_{{\rm Si}\,{\sc iii}}/N_{{\rm Si}\,{\sc iv}} \simeq 1$
suggests that 
$\log \bar{U} \simeq -2.2$, i.e. $\bar{U} \simeq 6.3\times10^{-3}$.
However, the mesoturbulent solution yields a
2.5 times higher value for the mean ionization parameter, i.e. 
$\bar{U} \simeq 1.6\times10^{-2}$
(for more details, see Levshakov, Agafonova \& Kegel 2000).

It is worthwhile to conclude this section with a few remarks concerning
the VPF method. From a mathematical point of view the VPF is quite
consistent since it simply means the representation of the unknown radial
velocity distribution function by the sum of several Voigt functions.
In such a way we can very well reproduce the line profiles and, in
the case of unsaturated lines, 
calculate accurately the total column densities.
But if we
wish to go further and estimate physical parameters like kinetic
temperature, ionization and metallicity, the VPF may produce misleading results.
If reality corresponds to 
a model like that shown in Fig.~1, it is physically unjustified
to interpret each subcomponent as belonging to one distinguished cloud
since as shown in Fig.~1 the contribution to
any point within the line profile does not come from a single separate
area but from volume elements distributed over the whole absorbing
region.

\begin{figure*}
\vspace{5.0cm}
\hspace{0.7cm}\psfig{figure=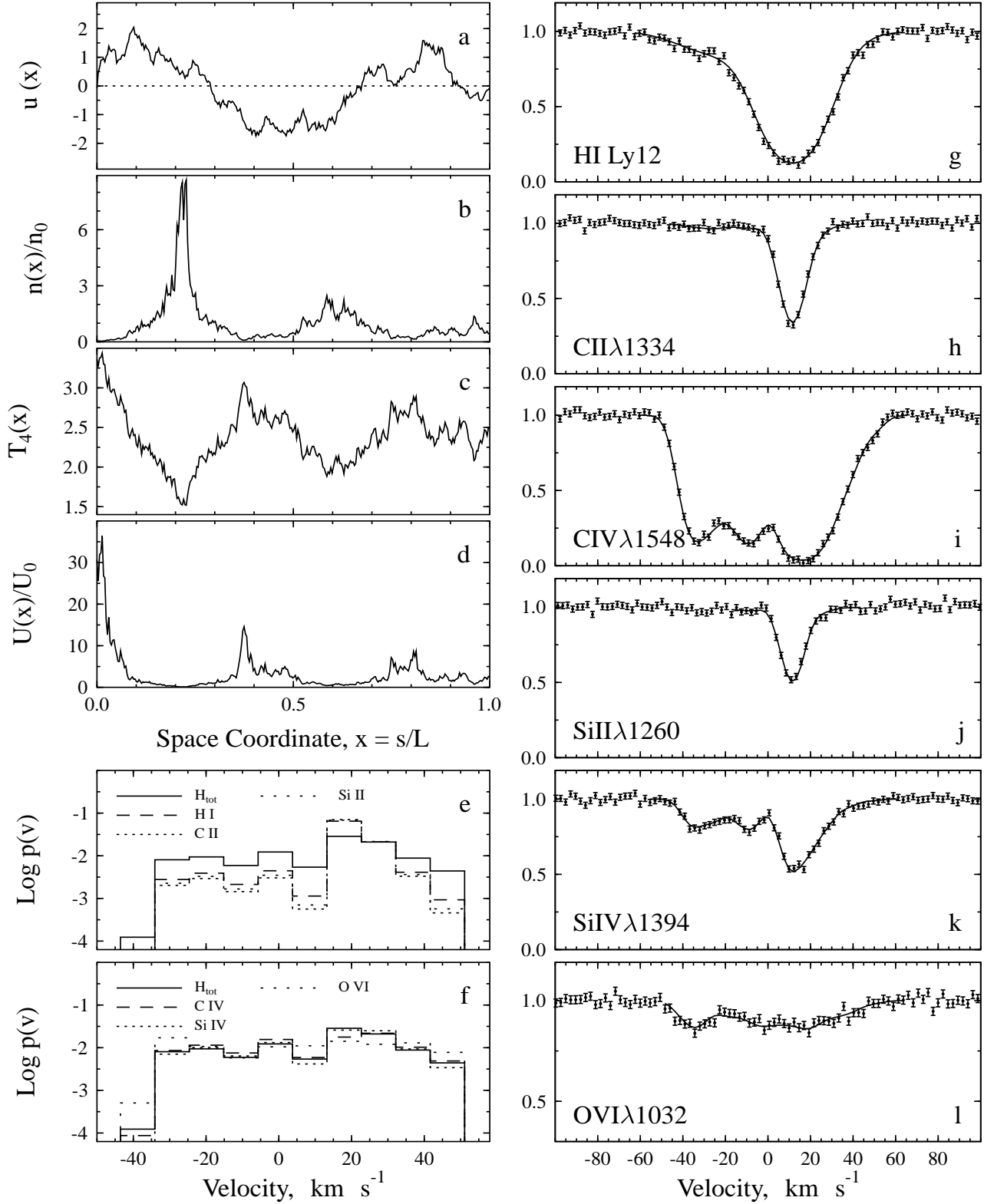,height=18.0cm,width=16.0cm}
\vspace{-1.8cm}
\caption[]{Model~A, Markovian fields~: 
the normalized velocity (a), density (b), kinetic temperature (c),
and ionization parameter (d) distributions vs space coordinate ($T$ is given
in units of $10^4$~K). The corresponding density-weighted distribution
functions for the radial velocities of the gas and the individual ions are 
shown in panels (e) and (f). Panels (g)~--(l) represent simulated spectra
convolved with an instrumental profile of FWHM = 7 km~s$^{-1}$ and added 
Gaussian noise with S/N = 50 (dots with error bars). The smooth curves show
the result of the MCI fitting 
}
\end{figure*}

\section{The Monte Carlo Inversion}

The Monte Carlo Inversion (MCI) of line profiles includes the estimation
of the physical parameters and simultaneously of three random functions
$n_{\rm H}(x)$, $T(x)$, and $v(x)$. For the equilibrium models of DS,
the kinetic temperature 
is determined from the energy balance equation in the optically thin
limit. In this case $T$ depends on 
the gas density only, 
the dependence being described approximately by a power law, i.e.
\begin{equation}
T \propto n_{\rm H}^{\gamma}\: ,\;\; {\rm with}\;\; \gamma < 0.
\label{eq:E13}
\end{equation}
For an individual realization the
remaining two random functions $n_{\rm H}(x)$ and 
$v(x)$ are determined as follows.

The hydrodynamic velocity and 
density fields are formed as a result of interference
of many independent and random factors -- 
infall and outflows, rotation, tidal flows, shock waves etc. 
In this case, we may consider the 
fluctuating amplitude of the velocity or the density
along the line of sight as a kind of Brownian motion which is mathematically
described by the so-called Markovian processes (e.g. Rytov et al. 1989).
In particular, in case of Gaussian fields the 
finite difference representation of the velocity as a Markovian process
has the form
\begin{equation}
v(x + \Delta x) = f_{\rm v}\,v(x) + \epsilon(x + \Delta x)\; ,
\label{eq:E14}
\end{equation}
where $f_{\rm v} = R_{\rm v}(\Delta x)/\sigma^2_{\rm v}$, 
$R_{\rm v}$ being the correlation between
the velocity values at points separated by a distance $\Delta x$, i.e.
$R_{\rm v} = \langle v(x + \Delta x)\,v(x) \rangle$, $\sigma_{\rm v}$ the 
velocity dispersion of bulk motions, and $\epsilon$ a random normal
variable with zero mean and dispersion
\begin{equation}
\sigma_{\epsilon,{\rm v}} = \sigma_{\rm v}\sqrt{1 - f^2_{\rm v}}\; .
\label{eq:E15}
\end{equation}

The density and velocity are 
in general related through the continuity equation. 
However, since we consider one component of the velocity
field only, the correlation between this component and the
density is less tight and in lowest order approximation
we may consider the two quantities to be statistically 
independent of each other.

Suppose further that the logarithmic density distribution is normal and that
the dimensionless variable
$y(x) = n_{\rm H}(x)/n_0$, with $n_0$ being the mean gas density 
has the second central moment $\sigma_{\rm y}$
characterizing the dispersion of the density field along the line of sight
[this definition implies that the expectation value $E(y) = 1$ for any $x$].
Then consider, by analogy with $v(x)$, 
a random function $\nu(x)$ 
with the following representation
\begin{equation}
\nu(x + \Delta x) = f_{\nu}\,\nu(x) + \epsilon(x + \Delta x)\; ,
\label{eq:E16}
\end{equation}
where $f_{\nu} = R_\nu(\Delta x)/\sigma^2_\nu$, 
$R_\nu$ being the correlation between
the values of $\nu$ at points separated by a distance $\Delta x$,
$\sigma_\nu$ the logarithmic density 
dispersion, and $\epsilon$ a random normal
variable with zero mean and dispersion
\begin{equation}
\sigma_{\epsilon,\nu} = \sigma_\nu\sqrt{1 - f^2_{\nu}}\; .
\label{eq:E17}
\end{equation}

Having defined $\nu(x)$, 
the distribution of the hydrogen density can be obtained
through Johnson's transformation (Kendall \& Stuart 1963)~:
\begin{equation}
n_{\rm H}(x) = n_0\,{\rm e}^{\nu(x) - \frac{1}{2}\sigma^2_\nu}\: ,
\label{eq:E18}
\end{equation}
with the relationship between 
the dispersions $\sigma_\nu$ and $\sigma_{\rm y}$
\begin{equation}
\sigma_\nu = \sqrt{\ln(1 + \sigma^2_{\rm y})}\: .
\label{eq:E19}
\end{equation}

With (\ref{eq:E18}), we can rewrite the expression (\ref{eq:E12}) for
$\tau_{{\rm a},i}(\lambda)$ in the form
\begin{eqnarray}
\lefteqn{
\tau_{{\rm a},i}(\lambda) = 
k_0 Z_{\rm a} N_0\,{\rm e}^{-\frac{1}{2}\sigma^2_\nu} } \nonumber \\ 
& & \times \int^1_0\,
{\rm e}^{\nu(x)}\,\Upsilon_{{\rm a},i}[U(x)]\,
\Phi_\lambda[\Delta\lambda_{\rm D}(x),v(x)]\,{\rm d}x .
\label{eq:E20}
\end{eqnarray}
where $N_0 = n_0\,L$ is the expectation value of the total hydrogen column density.

Equation (\ref{eq:E20}) can be simplified if we have a
fully resolved profile of an optically thin line. 
In this case the total ion column density along the given line of 
sight, $N_{{\rm a},i}$,
can be found directly from the observed profile (Paper~II), and thus we can
eliminate the unknown abundance from (\ref{eq:E20}). 
Integrating (\ref{eq:E20}) over $\lambda$ yields
\begin{equation}
Z_{\rm a} = \frac{N_{{\rm a},i}}{\bar{\Upsilon}_{{\rm a},i} N_0}\,
\label{eq:E23}
\end{equation}
where the mean density-weighted fractional ionization 
$\bar{\Upsilon}_{{\rm a},i}$ is
\begin{equation}
\bar{\Upsilon}_{{\rm a},i} =
{\rm e}^{-\frac{1}{2}\sigma^2_\nu} \int^1_0\,
{\rm e}^{\nu(x)}\,\Upsilon_{{\rm a},i}[U(x)]\,{\rm d}x\: .
\label{eq:E22}
\end{equation}
Inserting (\ref{eq:E23}) into (\ref{eq:E20}) leads to
\begin{eqnarray}
\lefteqn{
\tau_{{\rm a},i}(\lambda)\, = } \\
& & \frac{k_0\,N_{{\rm a},i}}{\bar{\Upsilon}_{{\rm a},i}}\,
{\rm e}^{-\frac{1}{2}\sigma^2_\nu}
\int^1_0 {\rm e}^{\nu(x)}\,\Upsilon_{{\rm a},i}[U(x)]\,
\Phi_\lambda[\Delta\lambda_{\rm D}(x),v(x)]\,{\rm d}x\,.\nonumber
\label{eq:E21}
\end{eqnarray}

Next, the obvious relationship
\begin{equation}
U(x) = \frac{n_{\rm ph}}{n_{\rm H}(x)} =
\frac{M J_0}{n_0}{\rm e}^{-\nu(x) + \frac{1}{2}\sigma^2_\nu}\: ,
\label{eq:E26}
\end{equation}
shows that 
the mean ionization parameter $U_0$ is given by the expression
\begin{equation}
\langle U(x) \rangle = U_0 = 
\frac{M J_0}{n_0}{\rm e}^{\sigma^2_\nu}
\label{eq:E27}
\end{equation}
(use again the Johnson transformation to evaluate the probability density
distribution of $U$).
Here $M$ is a multiplicative factor introduced by DS
in oder to express $n_{\rm ph}$ in terms of $J_0$.
For the photoionizing spectrum of Mathews \& Ferland, $M$
is $5.19\times10^{15}$ (in cgs units). 

\begin{figure}
\vspace{3.0cm}
\hspace{-1.7cm}\psfig{figure=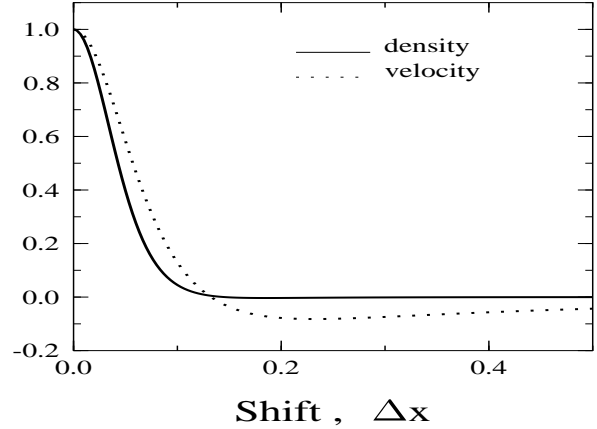,height=7.0cm,width=8.0cm}
\vspace{-4.2cm}
\caption[]{Model~B~: adopted correlation function of the velocity fluctuations
(dotted line) and corresponding correlation function of the
density contrast fluctuations (solid line)
}
\end{figure}

The result (\ref{eq:E27})
has quite interesting consequences. Namely, if the density is
completely homogeneous $(\sigma_\nu = 0)$ then $U_0 = M J_0/n_0$. But if the
density field is perturbed $(\sigma_\nu > 0)$,
then with the same mean density $n_0$ and the 
same background radiation flux $J_0$
one obtains a higher 
value of $U_0$ without any additional sources of ionization.
Intermittent regions of low and high ionization 
caused by the density fluctuations may occur in this case along 
a given line of sight.

Now we can describe our model in detail. It is
fully defined by specifying $N_0$ 
(the expectation value of the total column density),
the mean value of the ionization parameter $U_0$, the rms values of the
density  
and velocity fluctuations, $\sigma_{\rm y}$ and $\sigma_{\rm v}$, respectively,
the correlation coefficients $f_{\nu}$ and $f_{\rm v}$, and 
the abundances $Z_{\rm a}$
of the elements involved in the analysis.
All these parameters are components of the parameter vector $\theta$.
Additionally to these physical parameters, the density and velocity distributions
are to be known.
The continuous random functions $v(x)$ and $\nu(x)$ are represented in our 
computations by their 
values $v_j$ and $\nu_j$ sampled at the equally spaced spatial
points $x_j$. They are computed  with the 
relations (\ref{eq:E14}) and (\ref{eq:E16}) 
which correlate the values at 
neighboring points ($j = 1, 2, ... , {\rm k}$).
To estimate $\theta$, 
$\{v_j\}^{\rm k}_{j=1}$ and 
$\{y_j\}^{\rm k}_{j=1}$ from the absorption line profiles, we 
have to minimize the objective function
\begin{equation}
\chi^2 = \frac{1}{p}\sum^K_{\ell=1}\,\sum^{P_\ell}_{i=1}\,
\left[ {\cal F}^{\rm obs}_{\ell, \lambda_i} - 
{\cal F}^{\rm cal}_{\ell, \lambda_i}(\theta) \right]^2/\sigma^2_{\ell,i}\: .
\label{eq:E28}
\end{equation}
Here ${\cal F}^{\rm obs}_{\ell, \lambda_i}$ is the observed normalized intensity 
of spectral line $\ell$ according to eq.~(\ref{eq:E3}), 
and $\sigma_{\ell,i}$ is the experimental error 
within the $i$th pixel of the line profile.
${\cal F}^{\rm cal}_{\ell, \lambda_i}(\theta)$ is the simulated intensity 
of line $\ell$ at the same $i$th pixel
having wavelength $\lambda_i$. 
The total number of spectral lines involved in the
optimization procedure is labeled by $K$, and the total number of data points
$P = \sum^K_{\ell=1}\,P_\ell$, where $P_\ell$ is the 
number of data points for the $\ell$th line.
The number of degrees of freedom is labeled by $p$.

Let us go on to consider the correlation coefficients
$f_{\rm v}$ and $f_\nu$. 
As known, the Markovian processes 
have exponential
correlation functions~:\, 
$R(\Delta x) \propto \exp(-\Delta x/l)$, where $l$ is referred to as
the correlation
length and depends, in general, on the spatial scale. 
For the step size $\Delta x/l \ll 1$, 
the correlation coefficient $f$ is very close to unity.
The computational procedure is quite insensitive to the concrete values of $f$ 
because of the stochastic relations (\ref{eq:E14}) and (\ref{eq:E16}).
Therefore the value of $f$ cannot be estimated in the same
manner as other physical parameters.  It appears to be more suitable to fix a few sets
of $f_{\rm v}$ and $f_\nu$ and then to carry out the estimation of
other parameters with each of these sets. 
Hence, in reality
the parameter vector $\theta$ does not include components $f_{\rm v}$
and $f_\nu$. 

The computational scheme is similar to that described in Paper~III,
but in the present work we used instead of the classical 
Metropolis method the more efficient
generalized simulated annealing algorithm (Xiang et al. 1997).

Model calculations with different synthetic spectra have shown
that the minimization of $\chi^2$ in form of eq.~(\ref{eq:E28})
does not allow us to recover the physical
parameters with a sufficiently high accuracy (see also LTA).
To stabilize the estimation of $\theta$, we need to augment
$\chi^2$ by a regularization term (or a penalty function).
The choice of the penalty function
is rather heuristic and depends on the particular problem.
For instance, 
when we try to fit model spectra to real noisy data
it is reasonable to stop the minimization procedure at the value
of $\chi^2 \simeq 1$ in order to avoid fitting intensity fluctuations
caused by the noise. Another restriction stems from the 
restoring procedure for the density and velocity distributions.
If we use instead of the processes (\ref{eq:E14}) and
(\ref{eq:E16}) their normalized analogies~:\,
$v^\ast(x) = v(x)/\sigma_{\rm v}$ and
$\nu^\ast(x) = \nu(x)/\sigma_\nu$ with the dispersions $\sigma_{{\rm v}^\ast} =
\sigma_{\nu^\ast} = 1$, more stable estimations for $\sigma_{\rm v}$ and $\sigma_\nu$
are obtained. In addition, the process $\nu^\ast$ should be centered, i.e.
$\langle \nu^\ast \rangle = 0$\, (the center of the velocity distribution
is determined by the relative positions
of spectral lines and in principle
can differ from zero). Accounting for all these remarks,
the modified objective function is now written in the form
\begin{equation}
{\cal L} = |\chi^2 - 1| + \alpha (|\langle \tilde{\nu}^\ast \rangle| +
|\tilde{\sigma}_{\nu^\ast} - 1| +
|\tilde{\sigma}_{{\rm v}^\ast} - 1|)\; ,
\label{eq:E29}
\end{equation}
where $\alpha$ is a constant of order unity and
$\tilde{\nu}^\ast$, $\tilde{\sigma}_{\nu^\ast}$,
$\tilde{\sigma}_{{\rm v}^\ast}$ are the quantities
derived from the current configuration.
It follows that in the vicinity of the
global minimum we must have ${\cal L}_{\rm min} \approx 0$.

If the optimization of ${\cal L}$ gives $\langle v^\ast \rangle \neq 0$, the
estimated rms velocity  $ \sigma_{\rm v}$ is biased.  To correct its
value, the following relation  can be used
\begin{equation}
\sigma^{\rm corr}_{\rm v} = \sigma_{\rm v}\,\left(1 - 
{\rm k}\,\langle v^\ast \rangle^2 /
\sum^{\rm k}_{j=1} (v^\ast_j)^2 \right)^{1/2}\; .
\label{eq:E30}
\end{equation}

The practical implementation of the proposed computational procedure
for recovering the physical parameters from spectral lines of different
ions is described in the following section.

\section{Numerical Experiments}

\begin{table}
\centering
\caption{Model parameters used to generate synthetic spectra}
\label{tab1}
\begin{tabular}{cccccc}
\hline
\noalign{\smallskip}
model & $n_0$ & $J_0$ & $L$ & $\sigma_{\rm v}$ & $\sigma_{\rm y}$ \\
  & cm$^{-3}$ & ergs/s~cm$^2$~Hz & kpc & km/s \\
\noalign{\smallskip}
\hline
\noalign{\smallskip}
A & 0.001 & $1.5\,10^{-21}$ & 5.5 & 25 & 1.5 \\
B & 0.005 & $3.0\,10^{-21}$ & 1.5 & 20 & 0.8 \\
\noalign{\smallskip}
\hline
\end{tabular}
\end{table}

We tested the numerical procedure described above by inverting 
synthetic spectra
with known physical parameters and known underlying 
velocity-density distributions
along the line of sight. 
We have considered two models A and B with the physical
parameters listed in Table~1.

To calculate random fields several procedures can be
utilized. The most simple one 
has been realized in model~A~: 
the velocity and gas density fields were produced by means
of two independent Markovian processes according to eqs. 
(\ref{eq:E14}) and (\ref{eq:E16}) with
$f_{\rm v} = 0.995$, $f_\nu = 0.999$ 
and $\Delta x = 1/300$.
The fractional ionization for the different ions 
included in the analysis were then 
calculated using the model of DS. It is
worth emphasizing that this model may not necessarily capture 
the physics most correctly since it assumes a specific type of the 
ionizing radiation,
but since our purpose in this paper is to show the principal possibility 
of recovering the adopted physical characteristics of the absorbing region
from the synthetic spectra, the DS model was chosen because of its
transparency and easiness to compute. 
The synthetic spectra were then calculated according to eq.~(\ref{eq:E12})
with fixed metallicity for all ions 
${\rm Z} = 0.1\,{\rm Z}_\odot$ 
(solar abundances were taken from Grevesse 1984).
To complete the procedure the spectra were 
convolved with a Gaussian point-spread function having FWHM = 7 km~s$^{-1}$
and a Gaussian noise with S/N = 50 was added to each pixel
(dots with error bars in Fig.~2).

\begin{figure*}
\vspace{5.0cm}
\hspace{0.7cm}\psfig{figure=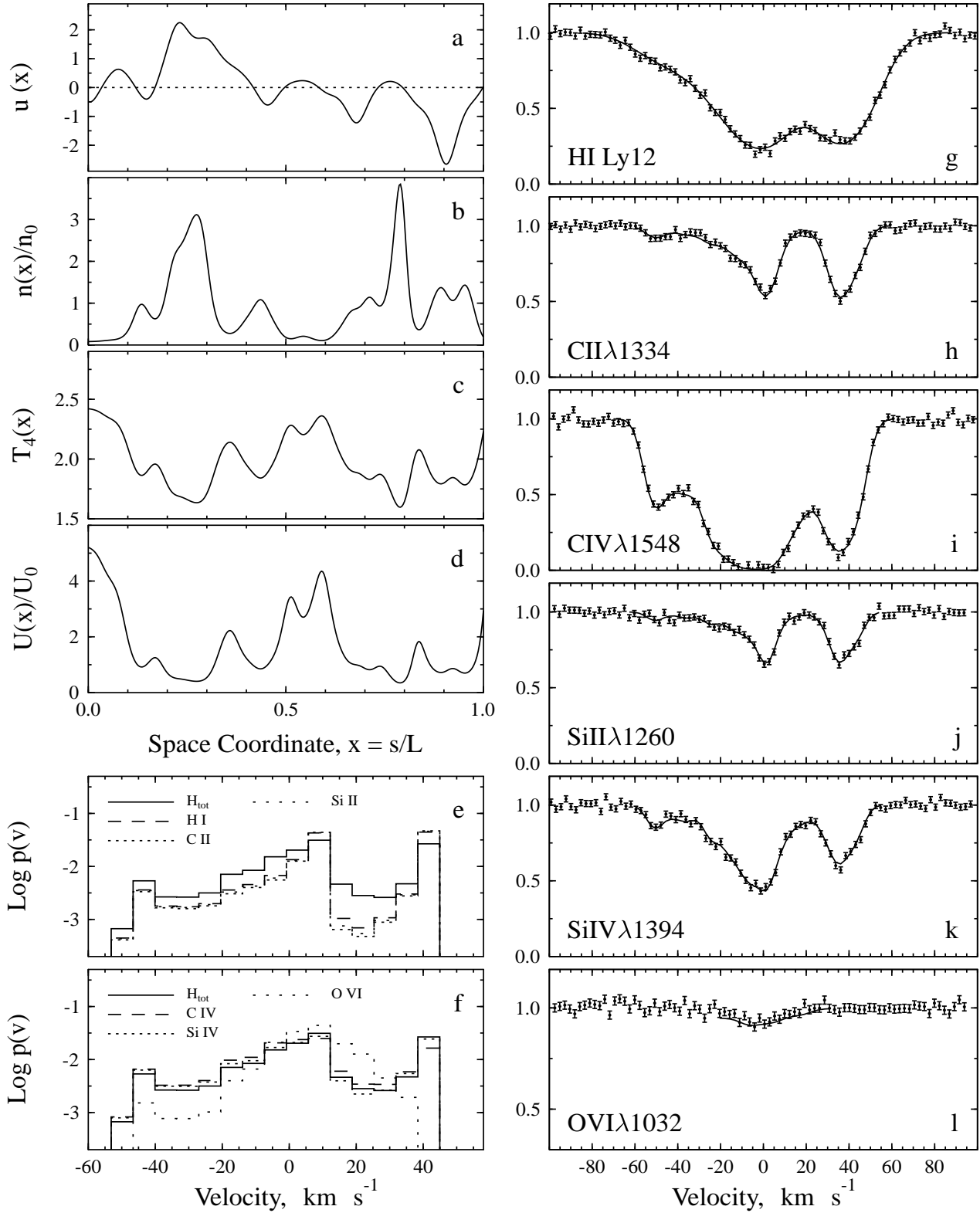,height=18.0cm,width=16.0cm}
\vspace{-1.8cm}
\caption[]{As Fig.~2, but for model~B, non-Markovian fields
}
\end{figure*}

Fig.~2 shows the hydrodynamic fields 
(normalized velocity $u = v/\sigma_{\rm v}$ and
gas density) together with the kinetic temperature 
(in units of $10^4$~K) and the ionization
parameter $U/U_0$ distributions along the line of sight 
(panels {\bf a}~--~{\bf d}). The histograms in 
panels {\bf e} and {\bf f} give the density-weighted 
velocity distributions of the different ions which 
determine the shapes of the spectral lines presented 
in turn in panels {\bf g}~--~{\bf l}
for the H\,{\sc i}~Ly12,  C\,{\sc ii}, C\,{\sc iv}, Si\,{\sc ii}, 
Si\,{\sc iv}, and O\,{\sc vi} lines, respectively. 
The Ly12 line was chosen for the analysis because 
it is unsaturated and, hence,
allows us to estimate the neutral hydrogen column density quite accurately. 
The choice of metallic ions was motivated by the
fact that their lines are formed in different parts of 
the adsorbing region~: 
C\,{\sc ii}, for example, needs denser gas and lower 
temperature to be produced 
(the area in the vicinity of $x \simeq 0.2$),
whereas  O\,{\sc vi} mainly is found in the very diffuse
and hot areas ($x \simeq 0.02, 0.38$ and 0.8). 
Such a diversity of lines in question
enables us to restore the underlying density and radial velocity 
distributions in more detail.

Model B differs from model A not only by the values of physical
parameters but by the method to generate random fields as well.
To produce the stochastic velocity and 
density contrast, $\delta(x)$, distributions, 
the corresponding power spectra were utilized.
The computational scheme was similar to that proposed by
Bi, Ge \& Fang (1995) but with another set of power spectra. 
We assume that the 1D power spectrum for the velocity
is given and then we calculate the 1D power spectrum for $\delta$.
In particular, for the normalized radial
velocity distribution, $u(x)$, the following 1D power spectrum was
chosen [it corresponds to the correlation function
$\cal{C}$$_{l,\varepsilon}(x)$ adopted in LTA]~:
\begin{equation}
S^{\rm 1D}_{\rm u}(k) =
\frac{k^{1-\varepsilon}\,{\rm e}^{-l\,k}}
{2\,\Gamma(2-\varepsilon)l^{\varepsilon -2}}\: ,
\label{eq:E31}
\end{equation}
where $\Gamma$ is the gamma function,  
$l$ and $\varepsilon$ are fixed constants 
($\varepsilon < 2$ and $l$ is in units of $L$). 

We can calculate the 1D power spectrum 
for the density contrast from its 3D spectrum which in turn 
is defined by the 3D power spectrum for the velocity~: 
$S^{\rm 3D}_\delta = k^2\,S^{\rm 3D}_{\rm u}$.
(This relation stems from the continuity equation 
when it is linearized in comoving coordinates).
The 3D power spectrum $S^{\rm 3D}_{\rm u}$
is related to its 1D spectrum via
(see Monin \& Yaglom 1975)~:
\begin{equation}
\frac{S^{\rm 1D}_{\rm u}(k)}{2 \pi k^2}  =
\int^\infty_k\,
\frac{S^{\rm 3D}_{\rm u}(k')}{k'}\,{\rm d}k'\: ,
\label{eq:E32}
\end{equation}
and thus
\begin{equation}
S^{\rm 3D}_{\rm u}(k) = \frac{1}{2 \pi k^2} 
\left[2\,S^{\rm 1D}_{\rm u}(k) - k\frac{{\rm d} S^{\rm 1D}_{\rm u}(k)}
{{\rm d} k}\right]\: .
\label{eq:E32a}
\end{equation}

The 1D power spectrum for the density 
contrast is then determined directly from its 3D spectrum
(Monin \& Yaglom)~:
\begin{equation}
S^{\rm 1D}_\delta(k)  = 2 \pi\,
\int^\infty_k\,
S^{\rm 3D}_\delta(k')\,k'\,{\rm d}k'\: .
\label{eq:E33}
\end{equation}

Given the power spectra $S^{\rm 1D}_{\rm u}(k)$ and
$S^{\rm 1D}_\delta(k)$,  which are the Fourier transforms of the
covariance functions of $u$ and $\delta$, both 
random fields $u(x)$ and $\delta(x)$ 
can be modeled using the moving-average method as described in LTA. 
In order to avoid the appearance of negative $n(x)$, the gas density itself
is assumed to have a log-normal distribution (see e.g. Bi, B\"orner \& Chu 1992; Bi \& Davidsen 1997). It is calculated from
$\delta$ according to
\begin{equation}
n(x)/n_0 = {\rm e}^{\delta(x) - \frac{1}{2}\langle \delta^2 \rangle}\: .
\label{eq:E34}
\end{equation}
The density contrast, as calculated from (\ref{eq:E34}), 
equals $\delta(x)$ in linear approximation.

The correlation functions of the processes 
$u(x)$ and $\delta(x)$ with the foregoing power spectra 
and with the adopted parameters $\varepsilon = 0.1$ and $l = 0.1$
are shown in Fig.~3. 
They differ significantly from the exponential
form corresponding to the pure Markovian process. 
So the objective to utilize these kind of stochastic
fields was to test whether we can approximate 
a finite realization of an arbitrary 
non-Markovian process by a Markovian one. 
The corresponding hydrodynamic fields and line profiles for model~B
are shown in Fig.~4 (notations are the same as in Fig.~2).

The recovering procedure aimed at the minimization of the objective function 
${\cal L}$, eq.~ (\ref{eq:E29}), consists of two
repeating steps~: firstly a random point in the physical parameter box is 
chosen and then the appropriate
optimal velocity and density configurations are estimated. 
(In our actual calculations we used instead of the
physical parameters $N_0$ and $U_0$ their reduced values 
$\hat{N}_0 = N_0\,{\rm e}^{-\frac{1}{2}\sigma^2_\nu}$   
and 
$\hat{U}_0 = U_0\,{\rm e}^{-\frac{1}{2}\sigma^2_\nu}$. 
This was made to achieve better stability of the results).

In both steps the stochastic acceptance rule was employed. 
The move in the parameter space or 
in the velocity-density configuration
space is accepted if it leads to a lower current value of ${\cal L}$. 
If ${\cal L}$ jumps up, then the move might be accepted
according to an acceptance probability $p_{\rm a}$.
Two types of the acceptance probability were
used in the procedure~: 
the standard Boltzmann-type statistics as utilized in the Metropolis algorithm
\begin{equation}
p_{\rm a} = {\rm e}^{-\Delta {\cal L}/{\cal T}}\: ,
\label{eq:E35}
\end{equation}
and the Tsallis statistics 
being the generalization of the Boltzmann distribution
(e.g. Xiang et al. 1997)
\begin{equation}
p_{\rm a} = [1 - (1 - q_{\rm a})\Delta 
{\cal L}/{\cal T}]^{1/(1 - q_{\rm a})}\: ,
\label{eq:E36}
\end{equation}
where $q_{\rm a}$ is a fixed constant and ${\cal T}$ is the so-called 
annealing parameter (`artificial temperature').

\begin{table*}
\centering
\caption{Physical parameters derived by the MCI method (rows 1-11)
from the synthetic spectra shown in Fig.~4. The first and the last row
lists, respectively, the adopted values (model~B) and
the median estimations}
\label{tab2}
\begin{tabular}{lccccccccrcc}
\hline
\noalign{\smallskip}
 & $\hat{U}_0$ & $\hat{N}_0$ & $\sigma_{\rm v}$ & $\sigma_{\rm y}$ &
$Z_{\rm C}$ & $Z_{\rm Si}$ & $Z_{\rm O}$ & $\chi^2$ &
$\langle \tilde{\nu}^\ast \rangle$ &
$\tilde{\sigma}_{\nu^\ast}$ &
$\tilde{\sigma}_{{\rm v}^\ast}$ \\
 & & cm$^{-2}$ & km/s \\
\noalign{\smallskip}
\hline
\noalign{\smallskip}
 &4.426e-3&1.583e19&20.0&0.8&4.900e-5&3.550e-6&8.100e-5\\
\noalign{\medskip}
1&4.392e-3&1.819e19&20.04&0.722&4.839e-5&3.488e-6&5.318e-5&1.057&-0.008&1.020&1.025\\
2&3.871e-3&1.744e19&22.95&0.729&4.835e-5&3.558e-6&8.409e-5&1.031&0.033&1.017&1.006\\
3&5.532e-3&1.580e19&24.57&1.215&4.853e-5&3.540e-6&4.516e-5&1.046&-0.028&1.009&0.937\\
4&3.878e-3&1.727e19&21.08&0.803&4.974e-5&3.602e-6&8.336e-5&1.095&4.0e-4&0.993&1.009\\
5&4.081e-3&1.837e19&21.97&0.695&4.936e-5&3.490e-6&7.794e-5&1.100&-0.013&1.005&0.986\\
6&3.104e-3&1.964e19&22.92&0.447&4.919e-5&3.456e-6&1.406e-4&1.095&-2.0e-4&1.012&0.966\\
7&5.683e-3&1.490e19&20.98&1.419&4.865e-5&3.628e-6&5.097e-5&1.103&-0.019&0.991&1.001\\
8&4.083e-3&1.780e19&19.56&1.206&4.988e-5&3.458e-6&7.769e-5&1.093&-0.068&0.983&1.096\\
9&4.520e-3&1.546e19&22.77&1.012&4.840e-5&3.701e-6&7.674e-5&1.151&-0.004&1.002&0.991\\
10&4.980e-3&1.630e19&26.05&0.998&4.913e-5&3.516e-6&7.615e-5&1.155&0.008&1.014&0.994\\
11&3.848e-3&1.821e19&20.28&0.679&4.961e-5&3.535e-6&8.471e-5&1.090&-0.013&1.000&1.017\\
\noalign{\medskip}
 &4.083e-3&1.727e19&21.97&0.803&4.913e-5&3.535e-6&7.674e-5 \\
\noalign{\smallskip}
\hline
\end{tabular}
\end{table*}

In general the acceptance after Tsallis leads to quicker
convergency, but it depends strongly on the value of 
$q_{\rm a}$ and requires a few initial
trial computations to adjust this parameter.
The annealing parameter ${\cal T}$
should decrease with the simulation 
time to ensure the
convergence to the global minimum. We used the `cooling' rate of type
${\cal T} = {\cal T}_0/(1 + N_{\rm steps})^q$,
with the initial value ${\cal T}_0$ and 
the power index $q$ having different values for both procedure steps
(i.e. the move in the parameter space
and the move in the density-velocity configuration space). 
Their values were adjusted in numerous
numerical experiments with the aim to reach better stability 
and quicker convergency. The results presented below were obtained 
with ${\cal T}^{\rm par}_0 = 2.0$, 
${\cal T}^{\rm d-v}_0 = \ln\,(\chi^2_0)$, 
$q_{\rm par} = 0.5$, 
$q_{\rm d-v} = 2$  and 
$q_{\rm a} = -1$ (here $\chi^2_0$ is the $\chi^2$ value for the
initial step in the parameter space).
The visiting distribution (a local search distribution providing trial values) 
was  Gaussian in all cases
with the fixed dispersion of 0.1 (in dimensionless units).

\begin{figure*}
\vspace{5.0cm}
\hspace{0.7cm}\psfig{figure=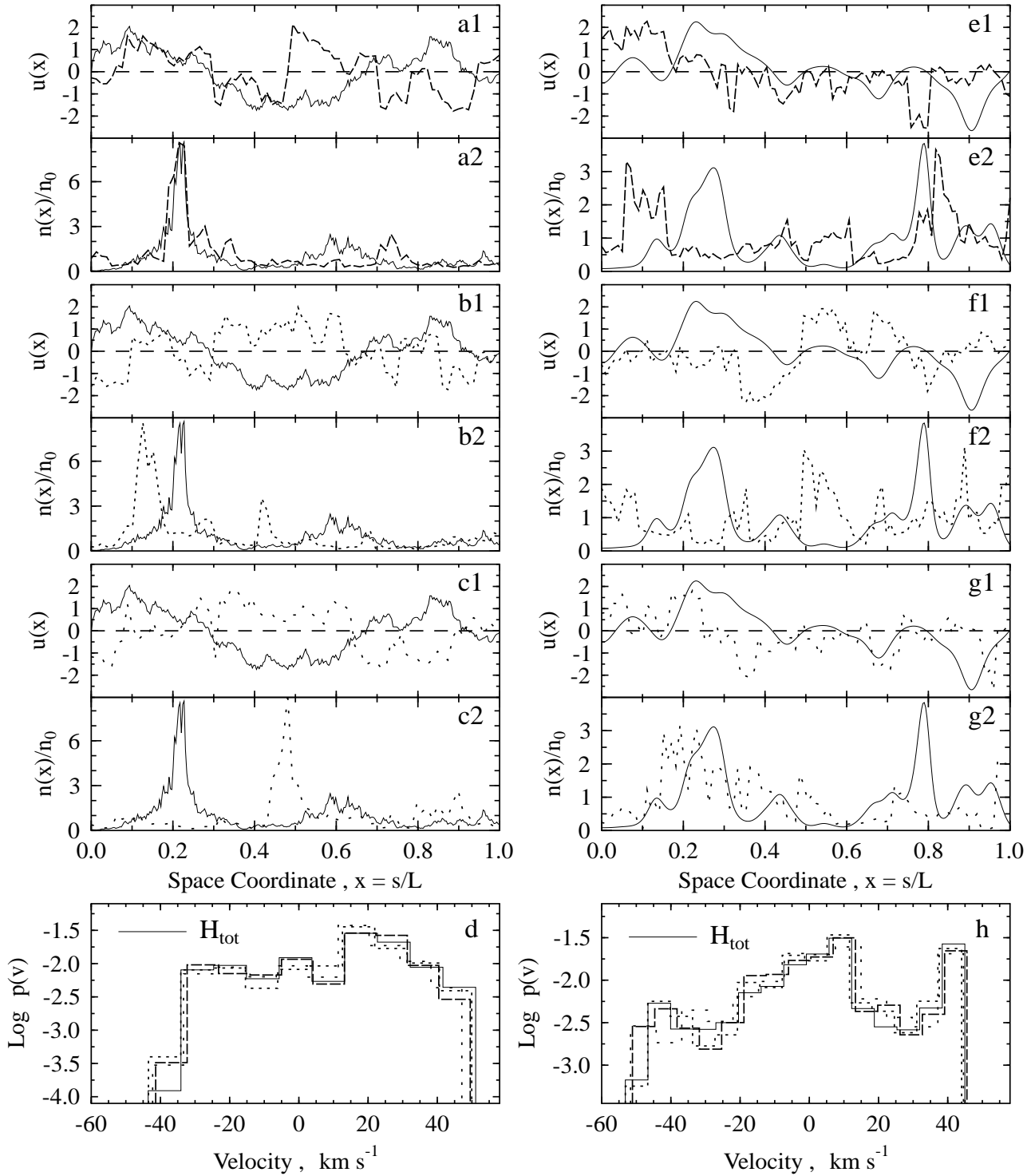,height=18.0cm,width=16.0cm}
\vspace{-1.8cm}
\caption[]{MCI reconstruction of the velocity and density fields for 
three different runs (a1,a2), (b1,b2), and (c1,c2) for model~A, and
(e1,e2), (f1,f2), and (g1,g2) for model~B (different types of dashed 
and dotted lines). The adopted distributions are shown by solid lines.
Panels (d) and (h) represent the corresponding density-weighted velocity
distribution functions (solid line histograms are adopted distributions,
whereas dashed and two types dotted lines are the MCI solutions)
}
\end{figure*}

The results of the computations are 
shown in Fig.~2 (model~A, Markovian fields) and in Fig.~4 
(model~B, non-Markovian fields) by solid lines (panels {\bf g}~--~{\bf l}). 
For both cases the step size was $\Delta x = 1/100$ and 
the correlation coefficients 
$f_{\rm v} =  f_\nu = 0.95$.  
These values were chosen as the optimal ones after several 
trial runs with different $\Delta x$, $f_{\rm v}$ and $f_\nu$.

The MCI is a probabilistic procedure. 
It follows that one does not know in advance if the global
minimum of the objective function is reached in a single run.
Several runs with different initial values should be performed 
to obtain the parameter estimations and their confidence levels. 
To illustrate the probabilistic character of the MCI solutions,
Fig.~5 shows the results of three different runs for model~A
(left-side panels) and for model~B (right-side panels).
Solid lines in all panels represent adopted fields, whereas dashed
and two types dotted lines are the MCI solutions. All
recovered fields yield $\chi^2 \la 1$, implying that the absorption spectra
calculated in different 
runs are indistinguishable from each other. The example 
shown by solid lines in Fig.~2 or 4 (panels {\bf g}~--~{\bf l})
is just one of the solutions obtained for the corresponding model~A or B.
However, it is not possible to restore the exact pixel-to-pixel 
structure of the stochastic 
velocity and density patterns along the line of sight -- many configurations 
are possible with equal probability. But we see that all
these configurations have the same
(statistically indistinguishable)
density-weighted velocity distributions 
(panels {\bf d} and {\bf h}),
since as already mentioned above only these
distributions determine the observed
line shapes.
It is obvious that the quality of recovering depends crucially
on the diversity of ion profiles accounted for in the MCI analysis~: 
the more ions originating from regions
with different physical conditions are considered the 
higher reliability of the estimated parameters can be reached.

Table 2 lists the results of 11 runs for model~B
(model~A shows entirely analogous results). It is seen
that the accuracy of the estimated parameters is rather high~:
the median values do not differ from the adopted ones by more
than 10\%. 
As for the confidence levels, it is obvious that
much more runs are needed to calculate them with a sufficient
accuracy. We have not carried out such computations since 
in the present context the exact values of the confidence
levels are not of importance. But from the runs performed we
can make some qualitative conclusions.  
For instance, the metallicities are estimated
with very high accuracy (the error in ${\rm Z}_{\rm O}$
is explained by the weakness of the oxygen line), i.e.
the numerical procedure is quite sensitive to these parameters.
The dispersions of $\sigma_{\rm v}$ and $\sigma_{\rm y}$ show
significantly wider margins caused by the randomness of the fields
which they characterize.
The estimations of $\hat{U}_0$ and $\hat{N}_0$ have an accuracy
of about 20\% that can be considered as quite acceptable for these
physical parameters.
A higher accuracy can be obtained for the estimations of the
total column densities of the individual ions and for the mean
(density-weighted) kinetic temperatures
\begin{equation}
\langle T_{{\rm a},i} \rangle =
\frac{\int\,n_{{\rm a},i}(x)\,T(x)\,{\rm d}x}
{\int\,n_{{\rm a},i}(x)\,{\rm d}x}\: .
\label{eq:E37}
\end{equation}
These physical parameters are listed in Table~3 in the same order as
the results presented in Table~2.

To complete this section, we note that
the computational time of one run with the adopted set of the spectral lines
and with two 100 components vectors representing the velocity and density
random configurations takes about 10 hours 
on a Pentium~III processor with 500 MHz.   

\begin{table*}
\centering
\caption{Physical parameters derived by the MCI method (rows 1-11)
from the synthetic spectra shown in Fig.~4. The first and the last row
lists, respectively, the adopted values (model~B)
and the median estimations} 
\label{tab3}
\begin{tabular}{lcccccccccccc}
\hline
\noalign{\smallskip}
  &H\,{\sc i}&C\,{\sc ii}&C\,{\sc iv}&Si\,{\sc ii}&Si\,{\sc iv}&O\,{\sc vi}& 
H\,{\sc i}&C\,{\sc ii}&C\,{\sc iv}&Si\,{\sc ii}&Si\,{\sc iv}&O\,{\sc vi} \\
\noalign{\smallskip}
\hline
\noalign{\smallskip}
  &\multicolumn{6}{c}{{\it Column densities in} cm$^{-2}$} &
\multicolumn{6}{c}{{\it Mean kinetic temperatures in} $10^4$~K}\\
\noalign{\smallskip}
\hline
\noalign{\smallskip}
  &5.92e16&5.52e13&2.76e14&4.43e12&1.80e13&6.22e12&1.75&1.74&1.88&1.74&1.84&2.16\\
\noalign{\medskip}
1&5.95e16&5.24e13&2.86e14&4.33e12&1.77e13&5.14e12&1.74&1.72&1.93&1.73&1.87&2.14\\
2&6.07e16&5.61e13&2.65e14&4.55e12&1.75e13&5.90e12&1.74&1.72&1.89&1.73&1.84&2.14\\
3&5.96e16&5.52e13&2.77e14&4.45e12&1.77e13&6.05e12&1.75&1.73&1.91&1.73&1.85&2.32\\
4&5.85e16&5.50e13&2.79e14&4.40e12&1.79e13&5.55e12&1.74&1.72&1.89&1.73&1.85&2.14\\
5&6.06e16&5.68e13&2.86e14&4.43e12&1.76e13&6.64e12&1.74&1.72&1.91&1.72&1.86&2.12\\
6&5.98e16&5.56e13&2.82e14&4.33e12&1.78e13&6.13e12&1.74&1.73&1.87&1.74&1.84&1.97\\
7&5.96e16&5.51e13&2.74e14&4.53e12&1.78e13&6.30e12&1.73&1.72&1.92&1.72&1.86&2.31\\
8&5.96e16&5.65e13&2.87e14&4.34e12&1.77e13&6.07e12&1.75&1.74&1.89&1.74&1.85&2.21\\
9&5.91e16&5.47e13&2.55e14&4.61e12&1.75e13&6.48e12&1.73&1.72&1.89&1.73&1.84&2.24\\
10&5.97e16&5.58e13&2.79e14&4.42e13&1.76e13&8.73e12&1.74&1.73&1.90&1.73&1.85&2.26\\
11&5.88e16&5.55e13&2.83e14&4.37e13&1.78e13&5.98e12&1.75&1.75&1.90&1.74&1.85&2.10\\
\noalign{\medskip}
 &5.96e16&5.55e13&2.79e14&4.42e13&1.77e13&6.07e12&1.74&1.72&1.90&1.73&1.85&2.14\\
\noalign{\smallskip}
\hline
\end{tabular}
\end{table*}

\section{Conclusions}

We have developed a new method to solve the inverse problem in the
analysis of intergalactic (interstellar) hydrogen and metal lines
arising from clumpy stochastic media.
In the method, the random velocity and density configurations
along the line of sight are approximated by Markovian processes.
The global optimization method based upon simulated annealing is
then used to fit theoretical line profiles to a set of `observational'
data. 

The proposed procedure allows us to estimate the physical parameters
of the absorbing gas such as column densities, metal abundances,
mean (density-weighted) kinetic temperatures for each ion, and mean
ionization parameter together with the hydrodynamic characteristics --
the radial velocity dispersion and the dispersion of the density
fluctuations.

The computational scheme has been tested on a variety of synthetic
spectra that emulate modern observational data~:
the absorption lines of
H\,{\sc i}, C\,{\sc ii}, Si\,{\sc ii}, C\,{\sc iv},
Si\,{\sc iv}, and O\,{\sc vi} which are usually observed
in the Lyman limit systems ($N_{{\rm H}\,{\sc i}} \la 3\times10^{17}$ cm$^{-2}$).
The ionization structure of the absorbing region was calculated using the
standard photoionization model of Donahue \& Shull (1991) with a background
ionizing spectrum given by Mathews \& Ferland (1987).

The inversion procedure proved to be very effective and robust allowing us
to recover the physical parameters with reasonable
accuracy
albeit the structure of the random velocity and density fields cannot be
restored with a pixel-to-pixel conformity. However, the integral 
characteristics of these random fields, namely, the density-weighted
velocity distribution, can be estimated quite precisely. Thus we can
conclude that our procedure provides reliable results and can be applied
to the analysis of real data.

Note that while performing the inversion of absorption lines, one
has to take into account the following. All our computational tests
have been carried out under the assumption that the spectrum of the
ionizing radiation is known, i.e. we used the same Mathews \& Ferland
spectrum to generate `observational' data and to fit them with our
theoretical profiles. In reality, the characteristics of the ionizing
radiation are not known exactly. Therefore in real applications
several types of the background photoionizing spectra should be tried.
The problem how the computational results are affected by different types
of the background ionizing radiation will be studied in detail elsewhere.

The proposed method has been successfully applied to the analysis
of QSO high resolution spectral data with possible deuterium absorption
at $z_{\rm a} = 3.514$ towards APM~08279+5255 (Levshakov, Agafonova \&
Kegel 2000). It has been demonstrated that the blue-side asymmetry of
the hydrogen Ly$\alpha$ line can be explained quite naturally by an
asymmetric configuration of the velocity field only. The results obtained 
revealed a considerably lower neutral hydrogen column density
as compared with the VPF measurements performed by Molaro et al. (1999).
In contrast to Molaro et al., we have managed to fit simultaneously
all absorption lines observed in this system. These results can be
considered as encouraging and favor the application of the developed 
computational procedure to the analysis of other high quality
observational data.

\medskip\noindent
\begin{acknowledgements}
The authors are grateful to Ellison et al. for providing the
spectra of quasar APM~08279+5525.
SAL and IIA gratefully acknowledge the hospitality of the 
University of Frankfurt/Main and 
the European Southern Observatory (Garching) where this work was performed.
This work was supported by the Deutsche Forschungsgemeinschaft and by the
RFBR grant No.~00-02-16007.
\end{acknowledgements}


\begin{thebibliography}{}

\bibitem[1992]{bib}Bi, H. G., B\"orner, G., Chu, Y. 1992, A\&A, 266, 1

\bibitem[1995]{bi}Bi, H., Ge, J., Fang, L.-Z. 1995, ApJ, 452, 90

\bibitem[1997]{bi}Bi, H., Ge, J., Davidsen, A. F. 1997, ApJ, 479, 523

\bibitem[1998a]{b1}Burles, S., Tytler, D. 1998a, ApJ, 499, 699

\bibitem[1998b]{b2}Burles, S., Tytler, D. 1998b, ApJ, 507, 732

\bibitem[1991]{ds}Donahue, M., Shull, J. M. 1991, ApJ, 383, 511 (DS)

\bibitem[1999]{el}Ellison, S. L., Lewis, G. F., Pettini, M.,
Sargent, W. L. W., Chaffee, F. H., Foltz, C. B., Rauch, M.,
Irwin, M. J.  1999, PASP, 111, 946

\bibitem[1995]{fr}Ferland, G. J. 1995, HAZY, A Brief Introduction to CLOUDY
(Univ. Kentucky, Physics Dept. Internal Report)

\bibitem[1998]{ghs}Giovanelli, R., Haynes, M. P., Salzer, J. J.,
Wegner, G., Da Costa, L. N., Freudling, W. 1998, AJ, 116, 2632

\bibitem[1998]{gr}Gramann, M. 1998, ApJ, 493, 28

\bibitem[1984]{ge}Grevesse, N. 1984, Phys. Scripta, T8, 49


\bibitem[1963]{ks}Kendall, M. G., Stuart, A. 1963, The advance
theory of statistics (Griffin \& Co., London)

\bibitem[1999]{kt}Kirkman, D., Tytler, D. 1999, ApJ, 512, L5

\bibitem[1997]{lk}Levshakov, S. A., Kegel, W. H. 1997, MNRAS, 288, 787 (Paper~I)

\bibitem[1997]{lkm}Levshakov, S. A., Kegel, W. H., Mazets, I. E. 1997,
MNRAS, 288, 802 (Paper~II)

\bibitem[1998]{lk2}Levshakov, S. A., Kegel, W. H. 1998, MNRAS, 301, 323

\bibitem[1998a]{l2}Levshakov, S. A., Kegel, W. H., Takahara, F. 1998a,
ApJ, 499, L1

\bibitem[1998b]{l3}Levshakov, S. A., Kegel, W. H., Takahara, F. 1998b,
A\&A, 336, L29

\bibitem[1999]{l4}Levshakov, S. A., Kegel, W. H., Takahara, F. 1999,
MNRAS, 302, 707 (Paper~III)

\bibitem[1999]{l5}Levshakov, S. A., Takahara, F., Agafonova, I. I. 1999,
ApJ, 517, 609 (LTA)

\bibitem[1999]{lk3}Levshakov, S. A., Kegel, W. H. 1999, in 
Proceedings of the XIXth Moriond, eds. F. Hammer, T. X. Thuan,
V. Cayatte, B. Guiderdoni and J. Tran Than Van, (Fronti\'eres, Paris), p.~431

\bibitem[2000]{l6}Levshakov, S. A., Agafonova, I. I., Kegel, W. H. 2000,
A\&A, 355, L1

\bibitem[2000]{l33}Levshakov, S. A., Tytler D., Burles S. 2000,
Astron. Astrophys. Trans., 19, in press (astro-ph/9812114)

\bibitem[1987]{mf}Mathews, W. D., Ferland, G. 1987, ApJ, 323, 456

\bibitem[1999]{mo}Molaro, P., Bonifacio, P., Centurion, M., Vladilo, G. 1999,
A\&A, 349, L13

\bibitem[1975]{my}Monin, A. S., Yaglom, A. M. 1975, Statistical Fluid Mechanics
(Cambridge, MIT)

\bibitem[1999]{nh1}Nusser, A., Haehnelt, M. 1999a, MNRAS, 303, 179 

\bibitem[1999]{nh2}Nusser, A., Haehnelt, M. 1999b, astro-ph/9906406

\bibitem[1997]{rhs}Rauch, M., Haehnelt, M., Steinmetz, M. 1997, ApJ, 481, 601

\bibitem[1989]{rkt}Rytov, S. M., Kravtsov, Yu. A., Tatarskii, V. I. 1989,
Principles of statistical radio physics (Springer, Berlin)

\bibitem[1996]{t}Tytler, D., Fan, X.-M., Burles, S. 1996, Nature, 381, 207

\bibitem[1997]{w}Watkins, R. 1997, MNRAS, 292, L59

\bibitem[1997]{xsfg}Xiang, Y., Sun, D. Y., Fan, W., Gong, X. G. 1997,
Phys. Lett. A, 233, 216


\end{thebibliography}
\end{document}